\documentstyle[12pt]{article}
\begin{document}
\begin{center}
{\Large\bf Supersymmetry of the
Nonstationary

Schr\"odinger
Equation and Time-Dependent

\vspace{2mm} Exactly Solvable Quantum Models}%
\footnote{Talk presented by B.F. Samsonov at the VIII International
Conference on "Symmetry Methods in Physics", Dubna, Russia, 28 July --
2 August, 1997.}

\vspace{3mm}
{\bf Boris F. Samsonov,} and {\bf L. A. Shekoyan}

{\small {\it  Tomsk State University, 36 Lenin Avenue, Tomsk, 634050 Russia,

E-mail: samsonov@phys.tsu.tomsk.su
}}
\end{center}

\begin{quotation}
\noindent {\small {\bf Abstract ---}
New exactly solvable time-dependent quantum models are obtained with the
help of the supersymmetric extension of the nonstationary Schr\"odinger
equation.}
\end{quotation}

\begin{center}{1. \ INTRODUCTION}\end{center}

An essential ingredient of the conventional supersymmetric quantum mechanics
(for reviews see \cite{SUSY})
is the well known Darboux transformation \cite{Darb} for the
stationary Schr\"odinger equation. This transformation permits us to
construct new exactly solvable stationary potentials from known ones.
Similar constructions may be developed for the time-dependent Schr\"odinger
equation \cite{BS}.

Our approach to  Darboux transformation is based on a general notion of
 transformation operator introduced by Delsart \cite{Delsart}.
In terms of this notion Darboux \cite{Darb} studied differential first
order transformation operators for the Sturm-Liouville problem. This is the reason in our opinion to call every differential transformation operator {\it Darboux transformation operator}. Different approach to Darboux transformation is exposed in the book

\cite{Matv}. It is worthwhile mentioning that our approach in contrast to that of Ref. \cite{Matv} leads to real potential differences. This property is crucial for constructing the supersymmetric extension of the nonstationary Schr\"odinger equation.

\begin{center}{2. \ FORMALISM}\end{center}

In this section we review briefly the basic constructions leading to the supersymmetry of the nonstationary Schr\"odinger equation established in Ref. \cite{BS}.

Consider two time-dependent Schr\"odinger equations
\begin{equation}
\label{sc1}\left( i\partial _t-h_0\right) \psi \left( x,t\right) =0,\quad
h_0=-\partial _x^2+V_0\left( x,t\right) ,
\end{equation}
\begin{equation}
\label{sc2}\left( i\partial _t-h_1\right) \varphi \left( x,t\right) =0,\quad
h_1=-\partial _x^2+V_1\left( x,t\right) .
\end{equation}
We assume the potential $V_0(x,t)$ and the solutions of Eq. (\ref{sc1}), called {\it the} {\it %
initial Schr\"o\-din\-ger equation,} to be known.
By definition transformation operator denoted by $L$
transforms solutions $\psi (x,t)$ into solutions
$\varphi (x,t)=L\psi (x,t)$. It is obvious that this condition is fulfilled if $L$ participates in the following intertwining relation:
$
L\left( i\partial _t-h_0\right) =\left( i\partial _t-h_1\right) L.
$
In the simplest case of a first order differential operator $L$ this
equation can readily be solved with respect to operator $L$ and potential
difference $A(x,t)$
$$ L=L_1(t)[-u_x(x,t)/u(x,t)+\partial _x],\quad
L_1\left( t\right) =\exp \left[ 2\int \mathop{\rm d}t \mathop{\rm
Im}\left( \log u\right) _{xx}\right] \ ,  $$
$$ A(x,t)=V_1(x,t)-V_0(x,t)=-[\log \left| u(x,t)\right| ^2]_{xx}\ .  $$
 Note that the operator $L$ and the new
potential $V_1(x,t)$ are completely defined by a function $u(x,t)$
called {\it the transformation function}.  This function is a
particular solution to the initial Schr\"o\-din\-ger equation
(\ref{sc1}) subject to the condition
$ \left( \log u/\overline u\right) _{xxx}=0 $
called {\it the reality condition of the new potential}.

Operator $ L^{+}=-L_1(t)[\overline u_x(x,t)/\overline u(x,t)+\partial _x]$  which is Laplace adjoint to $L$
realizes the transformation in the inverse direction, i.e. the
transformation from the solutions of Eq. (\ref{sc2}) to the ones of Eq. (\ref
{sc1}). The product $L^{+}L$ is a symmetry operator for Eq. (%
\ref{sc1}) and $LL^{+}$ is the similar one for Eq. (\ref{sc2}).

With the help of the transformation operators $L$ and $L^{+}$ we build up
the time-dependent nilpotent supercharge operators
\begin{equation}
\label{Q}Q=\left(
\begin{array}{cc}
0 & 0 \\
L & 0
\end{array}
\right) ,\quad
Q^{+}=\left(
\begin{array}{cc}
0 & L^{+} \\
0 & 0
\end{array}
\right)
\end{equation}
which commute with the Schr\"odinger super-operator $iI\partial _t-H,$
where $H={\rm diag}\{h_0,$ $h_1\}$ is the time-dependent
super-Hamiltonian and $I$ is the unit $2\times 2$ matrix.
In general,
the super-Hamiltonian is not integral of motion for the quantum
system guided by the matrix Schr\"odinger equation \begin{equation}
\label{mse}\left( iI\partial _t-H\right) \Psi \left( x,t\right) =0\ .
\end{equation}
Two-component function $\Psi \left( x,t\right) $ belongs to the linear
space defined over complex number field and
spanned by the basis $\Psi _{+}=\psi e_{+}$, $\Psi _{-}=L\psi e_{-}$ where $%
e_{+}=\left( 1,0\right) ^T$ and $e_{-}=\left( 0,1\right) ^T$. The sign
"{\sl T}" stands for the transposition.

The operators (\ref{Q}) are integrals of motion for Eq. (\ref{mse}). Using
the symmetry operators $L^{+}L$ and $LL^{+}$ we may construct other
integral of motion for this equation: $S={\rm diag}\left\{ L^+L,LL^+\right\}$.
The operators $Q$, $Q^{+}$ and $S$ realize the
well-known superalgebra
$sl(1/1)$
$$[Q,S]=[Q^+,S]=0,\quad \left\{ Q,Q^+\right\}=S-\alpha I$$
where instead of the Hamiltonian we
see other symmetry operator. In general, the operators
$S$, $Q$, and $Q^+$ depend on time and consequently we have obtained the
{\it time-dependent superalgebra}.

\newpage
\begin{center}{3. \ HARMONIC OSCILLATOR WITH A TIME-VARYING FREQUENCY}\end{center}

Consider the Hamiltonian
\begin{equation}
h_0=-\partial _x^2+\omega ^2(t)x^2\ .
\label{eq87}
\end{equation}
Variety of potentials we may obtain by the technique described above depends on variety of solutions of the initial Schr\"odinger equation suitable for use as transformation functions.
A wide class of solutions may be found with the help of {\it
R-separation of variables} method \cite{Miller} based on the orbits
structure of the symmetry algebra with respect to the adjoint
representation of corresponding group symmetry.  Symmetry algebra of
the Schr\"odinger equation with Hamiltonian (\ref{eq87}) is well-known
Schr\"odinger algebra $G_2$ \cite{Miller}.  The following
representation of this algebra is suitable for our purpose
$$K_1=a-a^+,\quad K_{-1}=-i(a+a^+),\quad K_0=i,$$
$$K_{-2}=-i(a+a^+)^2,\quad K_2=-i(a-a^+)^2,\quad K^0=-2\left[
a^2-(a^+)^2\right] , $$
$$ a=\varepsilon \partial _x-i\dot \varepsilon x/2, \quad
a^+=-\bar \varepsilon  \partial _x+i {\bar \varepsilon }x/2,\quad
aa^+-a^+a=1/4  $$ where $\varepsilon
=\varepsilon (t)$ is a (complex) solution to the classical equation of
motion for the oscillator $\ddot \varepsilon (t)+4\omega ^2(t)\varepsilon
(t)=0$.

Five orbits are known for this algebra which give four nonequivalent
solutions  to the Schr\"odinger equation in R-separated variables
with respect to transformations from the Schr\"odinger group. Below is
a summary of all suitable transformation functions and
corresponding potentials.

1). Two orbits with the representatives $J_1=K_1$ and $J_1=K_2$ :
$$u(x,t)=u_\lambda +u_{\bar \lambda }=
\gamma^{-1/2}\cosh \left( \frac {\nu x}{8\gamma }+
\mu \nu \frac {\delta }{32\gamma }\right)  $$
$$
\times \exp \left[ \frac {ix^2\dot \gamma }{4\gamma }-
\frac {i\mu x}{8\gamma }
+i(\nu ^2-\mu ^2)\frac {\delta }{64\gamma }\right] ,\quad
\lambda =-\mu -i\nu ,\quad
L_1(t)=\gamma =(\varepsilon +\overline \varepsilon )/2\ ,$$
$$
V_1(x,t)=\omega ^2(t)x^2-\frac {\nu ^2}{32\gamma ^2}
\cosh ^{-2}\left( \frac {\nu x}{8\gamma }+
\mu \nu \frac {\delta }{32\gamma }\right) .
$$

2). The orbit with representative $J_2=K_2-K_1$ :
$$\psi _\lambda (x,t)=\delta ^{-1/2}\exp \left( ix^2\frac {\dot \delta }{4\delta }
-ix\frac {\gamma }{2\delta ^1}+i\frac {\gamma ^3}{6\delta ^3}+
i\lambda \frac \gamma \delta \right)  $$
$$
\times Q\left( 2^{-1/2}\left( \frac x\delta -
\frac {\gamma ^2}{2\delta ^2}\right) -2^{2/3}\lambda \right)
$$
where $\gamma =\varepsilon +\bar \varepsilon ,\  i\delta =\varepsilon -\bar
\varepsilon $, $\lambda $ is a separation constant,
and $Q(z)$ is an {\it Airy} function defined by the equation:  $Q''(z)=zQ(z)$.
An exactly solvable potential is expressed in this case through the Airy
function $Ai(z)$. It is an easy exercise to show that a regular on full real axis
potential may be obtained with the help of the second order Darboux
transformation operator with transformation functions $\psi _\lambda $
and $\psi _{\overline \lambda }$.

3). The orbit with the representative $J_3=K_2-K_{-2}$ gives several classes of potentials. First, we may choose solutions which form a discrete basis in the Hilbert space of states \cite{Man} as transformation functions
$$
u _n(x,t)=N_n\gamma ^{-1/4}
\left( \frac {\overline \varepsilon }{\varepsilon }\right) ^{n/2+1/4}
\exp \left( \frac {2i\dot \gamma -1}{16\gamma }x^2\right)
He_n\left( \frac {x}{2\sqrt \gamma }\right) ,\
\gamma =\varepsilon \overline \varepsilon
$$
where $He_n(z)=2^{-n/2}H_n(z/\sqrt 2)$ are the Hermite polynomials.
The second order Darboux transformation with transformation functions  $u_n$ and $u_{n+1}$ produces potentials
$$
V_n(x,t)=\omega^2(t)x^2-\frac
{1}{2\gamma }\left[ \frac {J''_n(z)}{J_n(z)} -\left( \frac
{J'_n(z)}{J_n(z)}  \right) ^2-1 \right] , \quad z=x/(2\sqrt \gamma )\ , $$
$$J_n\left( z\right)
=\sum_{k=0}^n\frac{\Gamma \left( n+1\right) }{\Gamma \left( k+1\right)
}He_k^2\left( z\right) =kJ_{k-1}\left( z\right) +He_k^2\left( z\right) \ ,
 $$

Second, we may use a general solution to the quantization equation for the operator $J_3$
$$
\quad u(x,t)=\overline \varepsilon ^{-1/2}
\exp \left( \frac {2i\dot \gamma +1}{16\gamma }x^2  \right)
\left[ C+{\rm erf}\left( \frac {x}{2\sqrt {2\gamma }}  \right)   \right] \ .
$$
This leads to potentials which in the case of $\omega (t)={\rm const}$ reduce to the well-known isospectral potentials.
$$
V(x,t)=\omega ^2(t)x^2-\frac {1}{4\gamma }
\left[ 1-2zQ^{-1}(z)e^{-z^2/2}-2Q^{-2}(z)e^{-z^2}  \right] ,
$$
$$Q(z)=\sqrt {\frac {\pi }{2}}
\left[ C+{\rm erf}\left( \frac {z}{\sqrt 2}  \right)  \right] ,\quad
z=\frac {x}{2\sqrt \gamma } \ ,\quad  \left| C\right| >1\ .$$
Other cases are similar to these described in \cite{BS} for the free particle
Schr\"odinger equation so that be omitted here.

\begin{center}4. \ TIME DEPENDENT SINGULAR OSCILLATOR \end{center}

Consider now the following Hamiltonian:
$$h_0=-\partial
_x^2+\omega ^2\left( t\right) x^2+gx^{-2}\ .$$
 Symmetry algebra of the
Schr\"odinger equation with this Hamiltonian is $su(1.1)\sim sl(2,R)$.
We use the following representation for this algebra:
$$K_{+}=2\left[ \left( a^{+}\right) ^2-\overline{\varepsilon
}^2gx^{-2}\right] ,\quad K_{-}=2\left[ a^2-\varepsilon ^2gx^{-2}\right] ,
$$ $$K_0=\frac 12\left( K_{-}K_{+}-K_{+}K_{-}\right) =\frac 12\left[
K_{-},K_{+}\right]  \ .$$
Consider solutions of the Schr\"odinger equation which are
eigenstates of $K_0$:  $K_0\varphi _\lambda (x,$ $t)=\lambda \varphi
_\lambda (x,t).$
When $\lambda =n+k$, $n=0,1,2,\ldots $  we have a discrete basis of the Hilbert space
$$\varphi _n(x,t)=2^{1/2-3k}\sqrt{\frac{n!}{\Gamma \left( n+2k\right) }} \,
\gamma ^{-k}\left( \frac{\overline{\varepsilon }}\varepsilon \right) ^{n+k}x^{2k-1/2} $$
$$\times \exp \left[ i\frac{x^2\dot \gamma }{8\gamma }-\frac{x^2}{16\gamma }\right] L_n^{2k-1}\left( \frac{x^2}{8\gamma }\right) ,\
k=\frac {1}{2} +\frac {1}{4}\sqrt {1+4g}\ ,
\gamma =\varepsilon \overline \varepsilon \ .$$

To construct spontaneously broken supersymmetric model we need transformation  functions $u(x,t)$ such that neither $u(x,t)$ nor $u^{-1}(x,t)$ are not from the Hilbert space and $u(x,t)$ is nodeless for all real values of $t$ and $x>0$. These conditions a
re fulfilled for the functions
$$u_p(x,t)=\gamma ^{-k}\left( \frac{\overline{\varepsilon }}\varepsilon \right) ^{-p-k}x^{2k-1/2}
\exp \left[ i\frac{x^2\dot \gamma }{8\gamma }+\frac{x^2}{16\gamma }\right] L_p^{2k-1}\left( \frac{-x^2}{8\gamma }\right) \ ,$$
$$K_0u_p(x,t)=-(p+k)u_p(x,t) \ .$$
These transformation functions create the following exactly solvable family of potential differences $A(x,t)=\omega ^2(t)x^2+gx^{-2}-V_1(x,t)$:
$$A(x,t)=A_p(x,t)=\frac 1{4\gamma }-\frac{4k-1}{x^2}-\frac 18\left( \frac{xL_{p-1}^{2k}\left( z\right) }{\gamma L_p^{2k-1}\left( z\right) }\right) ^2 $$
$$+\frac{x^2L_{p-2}^{2k+1}\left( z\right) +4\gamma L_{p-1}^{2k}\left( z\right) }{8\gamma ^2L_p^{2k-1}\left( z\right) }\ ,\quad z=-\frac{x^2}{8\gamma }\ . $$
To construct a model with exact supersymmetry we need transformation functions $u(x,t)$ such that $u^{-1}(x,t)$ is square integrable on semiaxis $x\ge 0$ and satisfies the zero boundary condition at the origin for all values of $t$. The following solution
 of the Schr\"odinger equation may be chosen in this case:
$$u_p(x,t)=\gamma ^{k-1}\left( \frac{\overline{\varepsilon }}\varepsilon \right) ^{k-p-1}x^{3/2-2k}
\exp \left[ i\frac{x^2\dot \gamma }{8\gamma }+\frac{x^2}{16\gamma }\right] L_p^{1-2k}\left( \frac{-x^2}{8\gamma }\right) , $$
$$K_0u_p\left( x,t\right) =\left( k-p-1\right) u_p\left( x,t\right) . $$
It is not difficult to establish the possible values of $p$.
If $p$ is even it may takes the values
$p<2k-1$ and $p=[2k]+1,\  [2k]+3,\ \ldots $.
For odd $p$ values we may use only
$p=[2k],\ [2k]+2,\ldots $, where $[2k]\equiv {\rm entire}(2k)$.
For regular potential differences we obtain
$$A_p\left( x,t\right) =\frac 1{4\gamma }+
\frac{4k-3}{x^2}-\frac 12\left( \frac{xL_{p-1}^{2-2k}\left( z\right) }
{2\gamma L_p^{1-2k}\left( z\right) }\right) ^2
+\frac{x^2L_{p-2}^{3-2k}\left( z\right) +4\gamma L_{p-1}^{2-2k}
\left( z\right) }{8\gamma ^2L_p^{1-2k}\left( z\right) } \ .$$

\begin{center}
5. CONCLUSION
\end{center}

The supersymmetry of the time-dependent Schr\"odinger equation based on
the nonstationary Darboux transformation is very useful to obtain a wide
class of exactly solvable nonstationary quantum models.
With the help of the Darboux transformation operator we may obtain
solutions for transformed equations. In particular, if we know the
coherent states for the initial system then by applying to them the
Darboux transformation operator we obtain coherent states for transformed
quantum system \cite{CS}. The coherent states are known \cite{Man}
for the systems considered here. Next step is to obtain and investigate
coherent states for the transformed systems. Corresponding results will be presented elsewhere.

\begin{center}ACKNOWLEDGMENT\end{center}

This research has been partially supported by RFBR grant No 97-02-16279.
One of the authors (BFS) would like to thank the organizers of the
VIII International Conference "Symmetry Methods in Physics" for their kind
invitation to this meeting.

\end{document}